\documentclass[twocolumn,aps,preprint,letter,10pt]{revtex4}
\usepackage{graphicx,graphics,float}
\usepackage{dcolumn}
\usepackage{bm}
\usepackage{amsmath,amssymb}
\usepackage{mathrsfs}
\usepackage{xcolor}
\begin{document}
\title{Thermalization of linear Fermi systems}
\author{Jose Reslen}\email{josereslen@mail.uniatlantico.edu.co}
\affiliation{Coordinaci\'on de F\'{\i}sica, Universidad del Atl\'antico,
Carrera 30 N\'umero 8-49, Puerto Colombia.}%
%

\newcommand{\Keywords}[1]{\par\noindent{\small{\em Keywords\/}: #1}}

\begin{abstract} 
The issue of thermalization in open quantum systems is explored from the
perspective of fermion models with quadratic couplings and linear baths. Both
the thermodynamic state and the stationary solution of the Lindblad equation
are rendered as a matrix-product sequence following a reformulation in
terms of underlying algebras, allowing to characterize a family of stationary
solutions and determine the cases where they correspond to thermal states. This
characterization provides insight into the operational mechanisms that lead
the system to thermalization and their interplay with mechanisms that tend to
drive it out of thermal equilibrium.
\\
\Keywords{Thermalization, Open Quantum Systems, Matrix Product States.}
\end{abstract}
\maketitle
\section{Introduction}
The question of whether an quantum system evolves toward a thermal state keeps
being a topic of intense research
\cite{mori,dhar,reichental,huse,eisert,rigol,linden,enej,bojan}. The issue
encompass important aspects at a fundamental level, as for instance the
possibility of establishing operational paradigms that work akin to the
Fokker-Planck equation but at a wider scope, incorporating a diversity of
quantum effects.  In general terms, coupling to an environment has been
traditionally approached under the Markovian approximation \cite{huelga,vega},
giving rise to several developmental strategies.  In one of them, the effect
of the environment is traced out after both system and environment have been
studied perturbatively as a joint isolated system. This leads to a master
equation with identifiable physical mechanisms, but undesirable leaks in the
trace of the density matrix.  In other view, fundamental requirements like
trace conservation and positivity are guaranteed by considering Lindblad
generators, but then the connection of these baths with physical mechanisms
gets blurred.  In this latter scenario, it becomes important to know whether
it is possible to devise bath operators that drift the system precisely toward
its thermal state, so that the effect of the environment, which is conditioned
by its thermalization role, could be separated from the effect of the driving,
which is expected to couple much strongly and in that regard oust the system
from its thermal equilibrium. This subject is engaged here by establishing
when the solutions of the Gorini-Kossakowski-Sudarshan-Lindblad equation,
hereafter the Lindblad equation \cite{lindblad,gorini}, coincide with thermal
states.  In general, it is observed that if the baths coefficients satisfy
certain relations, the non-equilibrium state will evolve toward
infinity-temperature stationary states every time the Hamiltonian coefficients
form an irreducible matrix.  Additionally, it is possible for the
non-equilibrium state to converge toward thermal states of finite temperature,
but only when the Hamiltonian coefficients are diagonal. These results are
derived here in the framework of linear systems, also known as Gaussian- or
quadratic-systems due to the fact that the Hamiltonian comes about as a sum of
products of pairs of modes.  The emphasis on linearity intends to highlight
what seems to be the most consequential feature of these structures: All their
dynamics takes place in a subspace, making it possible to express and compare
different physical processes using the same formalism.

The work is organized as follows. Section \ref{s02191} covers the fundamental
notions and notation used throughout the work. Section \ref{s10021} introduces
the concept of second space, which is obtained from the space of Majorana
operators of the original problem following a third quantization recast
\cite{prozen}. Next, this second-space formalism is applied on the
thermodynamic state in section \ref{09063}, providing the foundation to
subsequently apply a reduction protocol that allows to decompose this state as
a matrix-product representation. Section \ref{s10022} summarizes the procedure
by which the Non-Equilibrium Stationary State (NESS), or the stationary
solution of the Lindblad equation, can be expressed in the second space as a
matrix-product succession. Sections \ref{s10026} and \ref{s10027} present two
theorems that describe a wide family of solutions of the time-independent
Lindblad equation. The prospect and implications of conforming these solutions
to thermal states are addressed in these sections too. To provide practical
context, numerical simulations of the so-found solutions are included in
section \ref{s10024}. Lastly, conclusions and some remarks are displayed in
section \ref{s10025}.
\section{The Hilbert space of non-interacting fermions subject to an
environment}
\label{s02191}
Let us consider the following single-body Hamiltonian 
\begin{gather}
\hat{H} = \sum_{j=1}^N \sum_{k=1}^N h_{j,k} \hat{c}_j^\dagger \hat{c}_{k},
\text{ } h_{j,k} = h_{k,j} \equiv \hat{h}.
\label{e01251}
\end{gather}
Mode operators satisfy standard fermionic rules $\{\hat{c}_j,\hat{c}_k^\dagger
\} = \delta_j^k$ and $\{\hat{c}_j,\hat{c}_k \} = 0$. The Hamiltonian
coefficients, $h_{j,k}$, are taken as real. Integer $N$ is the number of
single-body states that can be accessed by fermions. The single-particle
eigenenergies, $\epsilon_k$, and normalized eigenvectors, $\epsilon_{j,k}$,
are the respective solutions of the single-body eigenvalue equation
\begin{gather}
\hat{h} |\epsilon_k \rangle =  \epsilon_k |\epsilon_k \rangle \rightarrow
\sum_{l=1}^N h_{j,l} \epsilon_{l,k} = \epsilon_k \epsilon_{j,k}. 
\label{e02181}
\end{gather}
The many-body eigenvalues, $E$, and normalized eigenvectors, $|E\rangle$, are
the solutions of the system's Schrodinger equation 
\begin{gather}
\hat{H}| E \rangle = E | E \rangle.
\label{e02182}
\end{gather}
The eigenenergies of Hamiltonian (\ref{e01251}) are given in terms of the
single-body eigenenergies by
\begin{gather}
E_{n_1 n_2 ... n_N} = \sum_{j=1}^N \epsilon_j n_j.
\label{e01254}
\end{gather}
Integers $n_j$ are occupation numbers that can be either $0$ or $1$. The
normalized eigenstate associated with the above energy can be expressed thus
\begin{gather}
|E_{n_1 n_2 ... n_N} \rangle = \prod_{k=1}^N \left ( \hat{f}_k^{\dagger} \right )^{n_k} |0\rangle,
\label{e01261}
\end{gather}
so that
\begin{gather}
\hat{f}_k^{\dagger} = \sum_{j=1}^N \epsilon_{j,k} \hat{c}_j^\dagger. 
\label{e08033}
\end{gather}
These are the system's eigenmodes and follow standard fermionic
anticommutation relations.  State $|0\rangle$ describes a configuration with
no fermions. The total number of particles in the above state, $n$, can be
obtained from
\begin{gather}
n = \sum_{j=1}^N n_j.
\label{e02191}
\end{gather}
Since Hamiltonian (\ref{e01251}) preserves the total number of particles, $n$
is a valid quantum number and we can write $\hat{H}_n$ in case it be necessary
to indicate the number of fermions. However, the system's coupling to the
environment integrates spaces with different total number of particles and the
symmetry is scrambled. This makes it necessary to consider a
total-number-of-particle operator, $\hat{M}$, which can be defined as
\begin{gather}
\hat{M}= \sum_{j=1}^N \hat{m}_j, \text{ } \hat{m}_j = \hat{c}_j^\dagger
\hat{c}_j.
\label{e02192}
\end{gather}
When in contact with a bath of inverse temperature $\beta$ and chemical
potential $\mu$, it is hypothesized that the system reaches a state of
maximized entropy named the thermodynamic state. Such a state comes given by
\begin{gather}
\hat{\rho}_{th} = e^{-\beta (\hat{H} - \mu \hat{M})}/\Xi,
\label{e02194}
\end{gather}
where $\Xi$ is the gran-canonical partition function 
\begin{gather}
\Xi = \sum_{n=0}^N e^{\beta \mu n} Tr(e^{-\beta \hat{H}_n}).
\label{e02195}
\end{gather}
By definition $\hat{H}_0=0$. It can be shown that this function can be
expressed in terms of the Hamiltonian's eigenenergies
\begin{gather}
\Xi = \prod_{j=1}^N (1+e^{-\beta(\epsilon_j-\mu)}).
\label{e08032}
\end{gather}
Likewise, each eigenmode displays the following occupation weight
\begin{gather}
f_k = Tr (\hat{f}_k^\dagger \hat{f}_k \hat{\rho}_{th}) =
\frac{1}{e^{\beta(\epsilon_k-\mu)}+1},
\label{e08034}
\end{gather}
also known as the Fermi-Dirac distribution, while $\mu$ is also known as the
Fermi energy.

Another way how the system can reach a stationary state is through the
interaction with baths that induce a trace-preserving evolution. Let us then
consider a family of baths whose $n$'th element is given by
\begin{gather}
\hat{L}_n = \sum_{j=1}^N v_j^{(n)} \hat{c}_j + w_j^{(n)} \hat{c}_j^\dagger.
\label{e01262}
\end{gather}
Coefficients $v_j^{(n)}$ and $w_j^{(n)}$ are real constants that
determine the intensity of the contribution of destruction and creation
operators respectively.  The NESS,
$\hat{\rho}_{ss}$, is found as the solution of the time-independent Lindblad
equation: 
\begin{gather}
\frac{d \hat{\rho}_{ss}}{d t} = -i[\hat{H},\hat{\rho}_{ss}] + 
\sum_{n} 2 \hat{L}_n \hat{\rho}_{ss} \hat{L}_n^\dagger - 
\{ \hat{L}_n^\dagger \hat{L}_n, \hat{\rho}_{ss} \}=0.
\label{lindblad}
\end{gather}
Both the NESS as well as the thermodynamic state can be obtained as canonical
MPS-representations exploiting the parallels between the functionality of the
physical fermion-space and an alternative fermion-space with reduced
complexity. Such an alternative space is being introduced in the following
section.
\section{The second space}
\label{s10021}
Each mode operator can be written in terms of a pair of Majorana operators,
$\hat{\gamma}_k=\hat{\gamma}_k^\dagger$, in the next fashion
\begin{gather}
\hat{c}_j = \frac{\hat{\gamma}_{2j-1} + i \hat{\gamma}_{2j}}{2}, \text{ }
\hat{c}_j^\dagger = \frac{\hat{\gamma}_{2j-1} - i \hat{\gamma}_{2j}}{2}. 
\label{e02183}
\end{gather}
Majorana operators satisfy $\{\hat{\gamma}_j,\hat{\gamma}_k\} = 2 \delta_j^k$.
A key feature of this representation is that it is possible to establish an
exact parallel between an ordered string of Majorana operators and an element
of an alternative Fock space of fermions:
\begin{gather}
\hat{s} = \hat{\gamma}_1^{n_1} \cdots \hat{\gamma}_{2j-1}^{n_{2j-1}} (i
\hat{\gamma}_{2j})^{n_{2j}} \cdots (i \hat{\gamma}_{2N})^{n_{2N}}
\Leftrightarrow \nonumber \\
\left| n_1 \cdots n_{2j-1} n_{2j} \cdots n_{2N} \right ) = | s ).
\label{e02184}
\end{gather}
Just as in the previous section, the $n_j$'s are occupation numbers that can
be either $0$ or $1$. The above equivalence can be understood by observing how
a Fock state is defined \cite{schwabl}:
\begin{gather}
| s ) \equiv \left(\tilde{c}_1^\dagger \right)^{n_1} \cdots
\left(\tilde{c}_{2j-1}^\dagger \right)^{n_{2j-1}} \left(
\tilde{c}_{2j}^\dagger \right)^{n_{2j}} \cdots \left(\tilde{c}_{2N}^\dagger
\right)^{n_{2N}}|0). \nonumber
\end{gather}
As can be seen, the swapping of two neighbor operators in the above expression
prompts the same sign change than the swapping of corresponding operators in
$\hat{s}$ from (\ref{e02184}). In the previous equations curved kets and
tilded operators have been used to emphasize the fact that these elements
inhabit a Hilbert space that is different from the Hilbert space in which the
kets and operators used in section \ref{s02191} are defined, although both
spaces are governed by the same fermion algebra. Accordingly, operators with
a hat over them such as $\hat{s}$ and standard kets such as $| s \rangle$
should be considered as belonging to the {\it first space}, while operators
with a tilde over them such as $\tilde{s}$ and curved kets such as $| s )$
should be considered as belonging to the {\it second space}.  Notice that even
though these two sets describe the same problem, they do it from different
perspectives and therefore their physical interpretations are not
equivalent. In particular, only the first space admits a conventional
interpretation, i.e., association of kets with pure states and operators with
observables. To maintain a consistent notation, the same letter is used in
both spaces to represent the same physical concept. For example, the
Hamiltonian is $\hat{H}$ in the first space and $|H)$ in the second space.

It can be shown that the inner product in the second space has the
following equivalence
\begin{gather}
( s' | s ) \Leftrightarrow 2^{-N} Tr(\hat{s}'^\dagger \hat{s}).
\label{e02186}
\end{gather}
The trace operation above involves all the states in the basis of the first
space, not only those with an equal number of particles. Recall that $2^N$ is
the total number of many-body states that can be accessed by fermions in a
system of $N$ single-body states.

Operators on the second space can be used to replicate a number
of procedures in the first space. To see this, consider the following
expression:
\begin{gather}
i\hat{\gamma}_{2j} \hat{s} = i\hat{\gamma}_{2j} \hat{\gamma}_1^{n_1} \cdots (i
\hat{\gamma}_{2j})^{n_{2j}} \cdots (i \hat{\gamma}_{2N})^{n_{2N}}.
\label{e02201}
\end{gather}
The result depends on the value of $n_{2j}$. If $n_{2j}=0$ then
\begin{gather}
i\hat{\gamma}_{2j} \hat{s} = 
(-1)^{\sum_{k=1}^{2j-1} n_k}
\hat{\gamma}_1^{n_1} \cdots (i
\hat{\gamma}_{2j}) \cdots (i \hat{\gamma}_{2N})^{n_{2N}} \Leftrightarrow
\nonumber \\
(-1)^{\sum_{k=1}^{2j-1} n_k} \left| n_1 \cdots 1 \cdots n_{2N} \right ) =
\tilde{c}_{2j}^\dagger \left| n_1 \cdots 0 \cdots n_{2N} \right ). \nonumber
\end{gather}
The potential minus signs arises because in the process of taking
$i\hat{\gamma}_{2j}$ to its corresponding position in the ordered string a
minus sign ensues every time a couple of neighbor Majoranas are swapped.
If $n_{2j}=1$ then
\begin{gather}
i\hat{\gamma}_{2j} \hat{s} = 
-(-1)^{\sum_{k=1}^{2j-1} n_k}
\hat{\gamma}_1^{n_1} \cdots (i \hat{\gamma}_{2N})^{n_{2N}} \Leftrightarrow
\nonumber \\
-(-1)^{\sum_{k=1}^{2j-1} n_k} \left| n_1 \cdots 0 \cdots n_{2N} \right ) =
-\tilde{c}_{2j} \left| n_1 \cdots 1 \cdots n_{2N} \right ). \nonumber
\end{gather}
Both cases can be covered by a single expression independent of the value
of $n_{2j}$ in the next way
\begin{gather}
i\hat{\gamma}_{2j} \hat{s} \Leftrightarrow (-\tilde{c}_{2j} + \tilde{c}_{2j}^\dagger)
|s).
\label{e02204}
\end{gather}
\begin{table}
\begin{gather}
\begin{array}{|c|c|} \hline
\text{First space} & \text{Second space} \\ \hline
\hat{s} & | s ) \\ \hline
\hat{\gamma}_{2 j - 1} \hat{s}  & \left( \tilde{c}_{2 j - 1} + \tilde{c}_{2 j
- 1}^\dagger \right) | s ) \\ \hline
i \hat{\gamma}_{2 j}  \hat{s}  & \left( -\tilde{c}_{2 j} + \tilde{c}_{2
j}^\dagger \right) | s ) \\ \hline
\hat{s}  \hat{\gamma}_{2 j - 1}  & \left( -\tilde{c}_{2 j - 1} + \tilde{c}_{2
j - 1}^\dagger \right) (-1)^{\tilde M} | s ) \\ \hline
\hat{s}  i \hat{\gamma}_{2 j}  & \left( \tilde{c}_{2 j} + \tilde{c}_{2
j}^\dagger \right) (-1)^{\tilde M} | s ) \\ \hline
\end{array} \nonumber
\end{gather}
\caption{Both left and right multiplication of a string of ordered Majorana
operators by another operator have an equivalence on a fermionic Fock space.
Second space operator $\tilde {M}$ above has a meaning analogous to $\hat {M}$
in (\ref{e02192}).}
\label{t02201}
\end{table}
Following a similar analysis, the identities reported in table \ref{t02201}
can be obtained. These equivalences become fundamental in the process of
establishing the relations governing the physical state in the second space,
as described in the next two sections.
\section{Thermodynamic State in the Second Space}
\label{09063}
The thermodynamic state (\ref{e02194}) depends entirely on the following argument operator
\begin{gather}
\hat{R} = -\beta(\hat{H} - \mu \hat{M}) = \sum_{j=1}^N \sum_{k=1}^N R_{j,k}
\hat{c}_j^\dagger \hat{c}_k, 
\label{e03191}
\end{gather}
where
\begin{gather}
R_{j,k} = -\beta h_{j,k} + \beta \mu \delta_k^j,
\label{e08031}
\end{gather}
being $\delta_k^j$ the Kronecker delta.  It can be seen that
$R_{j,k}$ is symmetric because $h_{j,k}$ is symmetric too. In terms of
Majorana operators (\ref{e02183}) the above argument operator becomes
\begin{gather}
\hat{R} = \frac{i}{2} \sum_{j=1}^N \sum_{k=1}^N R_{j,k} \hat{\gamma}_{2j-1}
\hat{\gamma}_{2k} + \frac{1}{2} \sum_{j=1}^N R_{j,j}.
\label{e03192}
\end{gather}
This identity can also be expressed as
\begin{gather}
\hat{R} = \frac{i}{8} \sum_{j=1}^{2N} \sum_{k=1}^{2N} A_{j,k} \hat{\gamma}_{j}
\hat{\gamma}_{k} + A_0,
\label{e04091}
\end{gather}
so that
\begin{gather}
A_0 = \frac{1}{2} \sum_{j=1}^N R_{j,j},
\label{e04093}
\end{gather}
and
\begin{gather}
\hat{A} \equiv A_{j,k} = ((-1)^k - (-1)^j) R_{f(j),f(k)},
\label{e04092}
\end{gather}
where
\begin{gather}
f(l) = \frac{2l+1-(-1)^l}{4}.
\label{e10051}
\end{gather}
In this way, it is noticeable that matrix $\hat{A}$ is manifestly
antisymmetric $A_{j,k}=-A_{k,j}$. Following \cite{youla}, such a matrix can be
factorized as 
\begin{gather}
\hat{A} = \hat{U} \hat{\Lambda} \hat{U}^T,
\label{e04094}
\end{gather}
where $\hat{U}$ is orthonormal (real unitary) and 
{\small
\begin{gather}
\hat{\Lambda} = 
diag
\left (
\left (
\begin{array}{cc}
0 & \alpha_1 \\
-\alpha_1 & 0   \\
\end{array}
\right ), 
\left (
\begin{array}{cc}
0 & \alpha_2 \\
-\alpha_2 & 0   \\
\end{array}
\right ),
\dots,
\left (
\begin{array}{cc}
0 & \alpha_N \\
-\alpha_N & 0   \\
\end{array}
\right ) 
\right ).  \nonumber
\end{gather}
}
The $\alpha$'s are real coefficients. The thermodynamic state can
therefore be expressed as
\begin{gather}
\Xi \hat{\rho}_{th}(\vec{\xi}) = e^{A_0} e^{\frac{i}{4} \sum_{l=1}^{N} \alpha_l \hat{\xi}_{2l-1} \hat{\xi}_{2l}},
\label{e04095}
\end{gather}
where the $\xi$'s represent Majorana operators related to the $\gamma$'s via
transformation $\hat{U}^T$
\begin{gather}
\vec{\xi} =
\left [
\begin{array}{c}
\hat{\xi}_1    \\
\hat{\xi}_2    \\
\vdots    \\
\hat{\xi}_{2N}
\end{array}
\right ] =
\left [
\begin{array}{cccc}
U_{1,1} & U_{2,1}  & \dots   &  U_{2N,1}\\
 U_{1,2}& U_{2,2}  & \dots  &  U_{2N,2}\\
\vdots    & \vdots              &  \vdots &  \vdots \\
 U_{1,2N}& U_{2,2N} & \dots  & U_{2N,2N}
\end{array}
\right ]
\left [
\begin{array}{c}
\hat{\gamma}_{1}    \\
\hat{\gamma}_{2}    \\
\vdots      \\
\hat{\gamma}_{2N}
\end{array}
\right ].
\label{e04096}
\end{gather}
In order to reduce the thermodynamic state to a simpler form, next-neighbor
unitary operations are being applied on (\ref{e04095}). The key point in the
coming procedure is noticing that transformations applied on the whole state
can be tracked by the changes that they produce on the coefficients of the
matrix on (\ref{e04096}) as long as such transformations be linear. The
technique has been outlined in previous studies addressing topological
transitions in fermion systems \cite{reslen5} as well as quantum open
scenarios \cite{reslen6} and more recently under the effect of interaction
\cite{reslen7}. 

\begin{figure}[H]
\begin{center}
\includegraphics[width=0.5\textwidth,angle=0]{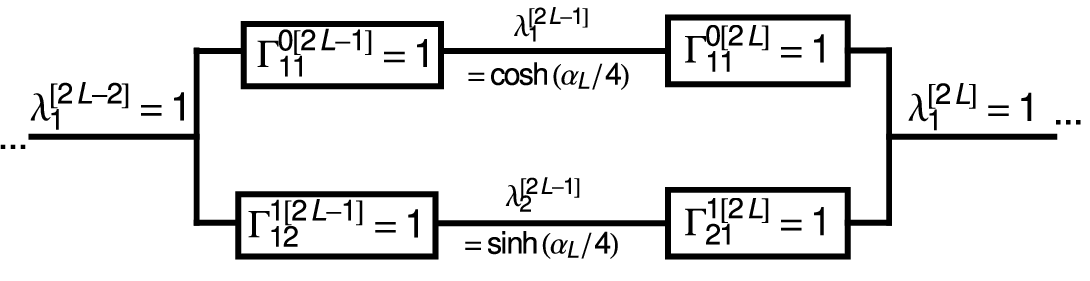}
\caption{Connection map of the tensorial representation of state
(\ref{e06036}).}
\label{fig1}
\end{center}
\end{figure}

Let us consider the following unitary operation
\begin{gather}
\hat{T}_{j,k}^{-1} = e^{\frac{\theta_{j,k}}{2} \hat{\gamma}_{j-1} \hat{\gamma}_j}.
\label{e05311}
\end{gather}
The effect of this transformation on the thermodynamic state has the following
equivalence
\begin{gather}
\hat{T}_{j,k}^{-1} \hat{\rho}_{th}(\vec{\xi}) \hat{T}_{j,k} =
\hat{\rho}_{th}(\hat{T}_{j,k}^{-1} \vec{\xi} \hat{T}_{j,k}).
\label{e05312}
\end{gather}
The result on a given mode is
{\small
\begin{gather}
\hat{T}_{j,k}^{-1} \hat{\xi}_k \hat{T}_{j,k} = U_{1,k} \hat{\gamma}_{1} + ...
U_{j-1,k}' \hat{\gamma}_{j-1} +  U_{j,k}' \hat{\gamma}_{j}  ... + U_{2N,k}
\hat{\gamma}_{2N}. \nonumber
\end{gather}
}
Only the coefficients of $\hat{\gamma}_{j-1}$ and $\hat{\gamma}_{j}$ are
affected. The updated coefficients read 
\begin{gather}
U_{j-1,k}' = U_{j-1,k} \cos \theta_{j,k} + U_{j,k} \sin \theta_{j,k}, \label{e05314} \\
U_{j,k}' = U_{j,k} \cos \theta_{j,k} - U_{j-1,k} \sin \theta_{j,k}. \label{e05313} 
\end{gather}
As a consequence, $U_{j,k}'$ can always be canceled by choosing
\begin{gather}
\tan \theta_{j,k} = \frac{U_{j,k}}{U_{j-1,k}}.
\label{e05315}
\end{gather}
In the first part, the protocol consists in applying a transformation
$\hat{T}_{N,1}$ on $\hat{\rho}_{th}$. The angle is chosen so as to cancel out
the term $U_{2N,1}$.  The effect of this operation on the thermodynamic state
is mirrored by the coefficients of matrix (\ref{e04096}) like follows 
\begin{gather}
\left [
\begin{array}{cccc}
U_{1,1} & \dots  & U_{2N-1,1}'   &  0 \\
 U_{1,2}& \dots  & U_{2N,-1,2}'  &  U_{2N,2}' \\
\vdots    & \vdots              &  \vdots &  \vdots \\
 U_{1,2N} &  \dots & U_{2N-1,2N}' & U_{2N,2N}'
\end{array}
\right ].
\label{e06032}
\end{gather}
Primed coefficients indicate updated factors, underlining the fact that the
transformation affects all the terms in the two rightmost columns. A second
transformation $\hat{T}_{N-1,1}$ is then applied to cancel out $U_{2N-1,1}'$.
The same cancellation procedure continues until only one coefficient at the
top row remains, at which point the matrix displays the following distribution
\begin{gather}
\left [
\begin{array}{cccc}
1 & 0  & \dots   &  0 \\
0 & U_{2,2}' &  \dots  & U_{2N,2}' \\
\vdots    & \vdots              &  \vdots &  \vdots \\
0 &  U_{2,2N}' &  \dots & U_{2N,2N}'
\end{array}
\right ].
\label{e06033}
\end{gather}
All the factors below $1$ must vanish because the matrix must be unitary. At
this point another set of transformations is applied to clear the second row
in the same way as with the first row, but this time the folding stops at the
second column lest the first row be unfolded. The clearing goes on in an
analogous way. When all the rows have been cleared the matrix is diagonal and
the reduced thermodynamic state is operationally equivalent to the original
one with the $\xi$'s replaced by $\gamma$'s. Moreover, the exponential can be
split because products of two Majorana operators commute
\begin{gather}
\Xi \hat{\rho}_{th}' = e^{A_0} e^{\frac{i}{4} \sum_{l=1}^{N}
\alpha_l \hat{\gamma}_{2l-1} \hat{\gamma}_{2l}} = e^{A_0} 
\prod_{l=1}^N e^{\frac{i}{4} \alpha_l \hat{\gamma}_{2l-1} \hat{\gamma}_{2l}}.
\label{e06034}
\end{gather}
The product argument can be expanded as
\begin{gather}
e^{\frac{i}{4} \alpha_l \hat{\gamma}_{2l-1} \hat{\gamma}_{2l}} = \cosh
\frac{\alpha_l}{4} + i \hat{\gamma}_{2l-1} \hat{\gamma}_{2l} \sinh
\frac{\alpha_l}{4}.
\label{e06035}
\end{gather}
Using this expression the reduced state can be written as a product state in
the second space along the lines of
\begin{gather}
|\Xi \rho_{th}') = e^{A_0} 
\bigotimes_{l=1}^N \left [ |0_{2l-1}0_{2l}) \cosh \frac{\alpha_l}{4} +
|1_{2l-1}1_{2l}) \sinh \frac{\alpha_l}{4} \right ].
\label{e06036}
\end{gather}
The subscript indicates the position in the Fock space where the occupation
apply. For instance, $|1)_{2l}$ means one fermion occupies the $2l$'th
single-body state.  Owing to the separability profile of state (\ref{e06036}),
it can be written as a product of matrices. In its canonical form, such
matrices keep a meaningful relation with the state's separability layout,
specifically, the elements of a canonical representation can be seen as the
coefficients of an expansion using local states plus Schmidt vectors as a
basis, like follows
\begin{gather}
| \rho ) = \sum_{\mu} \sum_{\nu}  \sum_{k=0}^1 \lambda_{\mu}^{L-1}
\Gamma_{\mu\nu}^{k L} \lambda_{\nu}^{L} |\mu ) |k ) |\nu ).
\label{e06191}
\end{gather}
State $|k)$ represents the local Fock basis at site $L$. Kets $|\mu )$ and
$|\nu )$ are Schmidt vectors spanning between the system's edges and the sites
to the left and right of $L$, respectively. As Schmidt vectors they must
satisfy $( \mu' | \mu ) = \delta_{\mu'}^\mu$ and $( \nu' | \nu ) =
\delta_{\nu'}^\nu$. Real factors $\lambda_\mu^{L-1}$ and $\lambda_\nu^L$ are
Schmidt coefficients associated to $|\mu )$ and $|\nu )$, respectively. Tensor
$\Gamma_{\mu \nu}^{k L}$ contains the superposition's coefficients. There is a
set of tensors $\lambda_\mu^{L-1}$, $\lambda_\nu^L$ and $\Gamma_{\mu \nu}^{k
L}$ for each $L$ and the group of these sets over $L=1,2,...,N$ form a
canonical decomposition of $| \rho )$. For a state like (\ref{e06036}) the
distribution of Schmidt vectors is discernible because the configuration is
separable. Figure \ref{fig1} depicts a set of coefficients that can be used to
represent state (\ref{e06036}).
Likewise, it can be determined using the equivalences of table \ref{t02201}
that the folding transformation (\ref{e05312}) has the following
correspondence in the second space
\begin{gather}
\hat{T}_{j,k}^{-1} \hat{\rho}_{th} \hat{T}_{j,k} \Leftrightarrow
\tilde{T}_{j,k}^{-1} | \rho_{th} ), 
\end{gather}
where
\begin{gather}
\tilde{T}_{j,k}^{-1} = e^{i \theta_{j,k} (-1)^{j} (
\tilde{c}_j^{\dagger} \tilde{c}_{j-1} + \tilde{c}_{j-1}^{\dagger} \tilde{c}_j)}.
\end{gather}
As a consequence, the thermodynamic state can be obtained in the second space
by reversely applying the inverses of the above transformations on the reduced
state described by figure \ref{fig1}. The thermodynamic state can therefore be
written as
\begin{gather}
|\Xi \rho_{th}) = \prod_{k=N-1}^1 \prod_{j=k+1}^N \tilde{T}_{j,k} |\Xi
\rho_{th}').
\label{e06221}
\end{gather}
The whole operation can be worked out in the matrix product representation
using the updating protocols of reference \cite{vidal} to calculate the effect
of neighbor unitary transformations on the tensor network. 
The grand-partition
function is encoded in the state's coefficients:
\begin{gather}
\Xi =  Tr(\Xi \hat{\rho}_{th}) \Leftrightarrow \Xi = 2^N (0|\Xi \rho_{th}).
\end{gather}
The value obtained in this way can be compared with the theoretical equivalent
(\ref{e08034}) for benchmarking. Notice that states (\ref{e06036}) and
(\ref{e06221}) have the same partition function because they are related by
unitary transformations. As a consequence, their respective overlap with $|0)$
must display identical values.  One can use this fact to get
\begin{gather}
\Xi =  2^N e^{A_0} \prod_{l=1}^N \cosh \frac{\alpha_l}{4}.
\label{e08041}
\end{gather}
Similarly, the eigenmode occupation of the thermodynamic state has the
following equivalent expression in the second space
\begin{gather}
f_k =  \frac{1}{2}\left(1 + 2^N \sum_{j=1}^N \sum_{i=1}^N
(-1)^{\theta(2j-2i-1)} \epsilon_{j,k} \epsilon_{i,k} \right . \nonumber \\
\left.   (0...1_{min(2j-1,2i)}...
0...1_{max(2j-1,2i)}... 0| \rho_{th} )\right),
\end{gather}
being $\theta(x)$ the Heaviside step function. Since one can build a common
spectrum for matrices $R_{j,k}$ and $h_{j,k}$ in (\ref{e08031}) it can be
assumed there exists a transformation that simultaneously diagonalize
$A_{j,k}$ and $h_{j,k}$ in such a way that the eigenmode occupations had
corresponding equivalences with the occupations of the reduced state
(\ref{e06036}) as follows
\begin{gather}
Tr (\hat{f}_k^\dagger \hat{f}_k \hat{\rho}_{th}) = Tr
(\hat{c}_k^\dagger \hat{c}_k \hat{\rho}'_{th}) \Leftrightarrow \nonumber \\
\frac{1}{2}\left( 1+2^N(0...1_{2k-1}1_{2k}...0|\rho'_{th}) \right).
\label{e08131}
\end{gather}
Using the coefficients in (\ref{e06036}) and partition function (\ref{e08041})
the mode occupation above yields
\begin{gather}
f_k = \frac{1}{2}\left(1 + \tanh \frac{\alpha_k}{4}\right) =
\frac{1}{e^{-\frac{\alpha_k}{2}}+1}.
\end{gather}
From a direct comparison with the Fermi-Dirac distribution in equation
(\ref{e08034}) the next equality is derived
\begin{gather}
-\frac{\alpha_k}{2}  = \beta(\epsilon_k - \mu).
\label{e08132}
\end{gather}
Knowledge of state (\ref{e06221}) and the grand-canonical partition function
(\ref{e08041}) provides an operational description that can be used to study
observables of interest.
\section{Non-Equilibrium Stationary State in the Second Space}
\label{s10022}
The time-independent Lindblad equation (\ref{lindblad}) goes over to the
second space in the following form
\begin{gather} 
\tilde{\mathscr{L}} | \rho_{ss} ) = 0.
\label{e07073} 
\end{gather} 
From the above expression, it can be readily appreciated that the NESS belongs
to the kernel of the Liouvillian.  Using the equivalences reported in table
\ref{t02201} to reformulate equation (\ref{lindblad}) in the second space it
can be found that the Liouvillian above is given by the next expression
\cite{reslen6}
{\scriptsize 
\begin{gather} 
\tilde{\mathscr{L}} = i\sum_{j=1}^{N} \sum_{k=1}^{N}  h_{j,k} \left (
\tilde{c}_{2 k}^\dagger \tilde{c}_{2 j - 1} + \tilde{c}_{2 j - 1}^\dagger
\tilde{c}_{2 k}\right ) + 2 \sum_n 
\label{e07071}  \\ 
(-B_{2j-1}^{(n)} \tilde{c}_{2j-1}^\dagger + B_{2j}^{(n)}
\tilde{c}_{2j}^\dagger)(B_{2k-1}^{(n)}  (\tilde{c}_{2k-1} +
\tilde{c}_{2k-1}^\dagger ) + B_{2k}^{(n)} (-\tilde{c}_{2k} +
\tilde{c}_{2k}^\dagger)) 
\nonumber \\ 
+ (B_{2j-1}^{(n)} \tilde{c}_{2j-1}^\dagger + B_{2j}^{(n)}
\tilde{c}_{2j}^\dagger)(B_{2k-1}^{(n)}(-\tilde{c}_{2k-1} +
\tilde{c}_{2k-1}^\dagger ) - B_{2k}^{(n)} (\tilde{c}_{2k} +
\tilde{c}_{2k}^\dagger)). \nonumber 
\end{gather} 
}
Factors $h_{j,k}$ correspond to the Hamiltonian coefficients of equation
(\ref{e01251}). In addition,
\begin{gather} 
B_{2j-1}^{(n)} = \frac{v_j^{(n)} + w_j^{(n)}}{2}, \hspace{0.25 cm}
B_{2j}^{(n)} = \frac{v_j^{(n)} - w_j^{(n)}}{2},
\end{gather} 
where $v_j^{(n)}$ and $w_j^{(n)}$ are the bath coefficients
introduced in equation (\ref{e01262}). It can be checked that a fully
occupied state is a right eigenstate of (\ref{e07071}): 
\begin{gather} 
\tilde{\mathscr{L}}|1...1) = L|1...1), \hspace{0.25 cm} 
L = -4\sum_n \sum_{j=1}^N B_{2j-1}^{(n)^2} + B_{2j}^{(n)^2}. \nonumber
\end{gather} 
The Liouvillian can also  be written in terms of Majorana operators defined in
the second space, $\tilde{\gamma}_j$, having the same functionalities that the
ones introduced in equation (\ref{e02183}):
\begin{gather}
\tilde{c}_j = \frac{\tilde{\gamma}_{2j-1} + i \tilde{\gamma}_{2j}}{2},
\hspace{0.25cm}
\tilde{c}_j^\dagger = \frac{\tilde{\gamma}_{2j-1} - i \tilde{\gamma}_{2j}}{2},
\hspace{0.25cm} \{ \tilde{\gamma}_j, \tilde{\gamma}_k \} = 2 \delta_j^k.
\nonumber
\end{gather}
Using these, the Liouvillian becomes
{\scriptsize
\begin{gather}
\tilde{\mathscr{L}} =  \sum_{j=1}^N \sum_{k=1}^N  \frac{h_{j,k}}{2}  
( \tilde{\gamma}_{4k} \tilde{\gamma}_{4j-3} + \tilde{\gamma}_{4j-2}
\tilde{\gamma}_{4k-1} ) +  \sum_n \label{e07072} \\
- B_{2j}^{(n)} B_{2k}^{(n)} \tilde{\gamma}_{4j} \tilde{\gamma}_{4k} 
- B_{2j}^{(n)} B_{2k}^{(n)} \tilde{\gamma}_{4j-1} \tilde{\gamma}_{4k-1} 
- B_{2j-1}^{(n)} B_{2k-1}^{(n)} \tilde{\gamma}_{4j-2} \tilde{\gamma}_{4k-2}
\nonumber \\
-B_{2j-1}^{(n)} B_{2k-1}^{(n)} \tilde{\gamma}_{4j-3} \tilde{\gamma}_{4k-3} 
+ 2 B_{2j-1}^{(n)} B_{2k}^{(n)} \tilde{\gamma}_{4j-2} \tilde{\gamma}_{4k} +
\nonumber \\
 2 B_{2j}^{(n)} B_{2k-1}^{(n)} \tilde{\gamma}_{4j-1} \tilde{\gamma}_{4k-3} 
+ 2 i \left( B_{2j-1}^{(n)} B_{2k-1}^{(n)} \tilde{\gamma}_{4j-2} \tilde{\gamma}_{4k-3} \right .
\nonumber \\
\left . + B_{2j}^{(n)} B_{2k}^{(n)} \tilde{\gamma}_{4j}
\tilde{\gamma}_{4k-1} + B_{2j-1}^{(n)} B_{2k}^{(n)} \tilde{\gamma}_{4j-3}
\tilde{\gamma}_{4k} + B_{2j-1}^{(n)} B_{2k}^{(n)} \tilde{\gamma}_{4j-2}
\tilde{\gamma}_{4k-1} \right), \nonumber
\end{gather}
}
or more simply
\begin{gather}
\tilde{\mathscr{L}} = \sum_{j=1}^{4N} \sum_{k=1}^{4N} \mathscr{L}_{j,k}
\tilde{\gamma}_{j} \tilde{\gamma}_{k}.
\label{e10052}
\end{gather}
The coefficients of $\mathscr{L}_{j,k}$ derive from a direct algebraic
correspondence between (\ref{e07072}) and (\ref{e10052}). Using the
anticommutation properties of Majorana fermions the above expression can be
written as an antisymmetric part plus a constant.

The NEES associated with (\ref{e07075}) can be found in the second space
following the procedure described in reference \cite{reslen6} in the context
of the open Kitaev chain subject to linear baths. Such a procedure is being
briefly outlined below to provide completeness. A more detailed develompment
can be consulted directly in reference \cite{reslen6}. The NESS is calculated
as the time evolution of a completely mixed state along infinity time. This
leads to the following expression
\begin{gather}
|\rho_{ss}) = \prod_{j=2N}^{1} \left ( \sum_{k=1}^{4N} R_{j,k} \tilde{\gamma}_k  \right) | 1\dots 1 ).
\label{e08133}
\end{gather}
The $R_{j,k}$'s are time-independent {\it complex} coefficients that depend
entirely on the Liouvillian factors $\mathscr{L}_{j,k}$. This state can be
folded exploiting the orthogonality properties of the coefficients and
following a reduction protocol where the imaginary parts are folded first and
the real parts later. Taking the inverse transformation, the NESS can be
written in the next way 
\begin{gather}
|\rho_{ss}) = z_0 \tilde{T}^{-1} | 1\dots 1 ).
\label{e08134}
\end{gather}
Coefficient $z_0$ is a normalization constant and 
\begin{gather}
\tilde{T} = \prod_{l=2N-1}^{1}  \prod_{m=2l+1}^{4N} \tilde{V}_{m,l}
\prod_{k=2l}^{4N} \tilde{U}_{k,l},
\end{gather}
being
\begin{gather}
\tilde{U}_{k,l} = e^{\frac{\theta_{k,l}}{2} \tilde{\gamma}_{k}
\tilde{\gamma}_{k-1}} \text{ and } \tilde{V}_{k,l} = e^{\frac{\phi_{k,l}}{2} \tilde{\gamma}_{k} \tilde{\gamma}_{k-1}}.
\end{gather}
Angles $\theta_{k,l}$ and $\phi_{k,l}$ are prescribed by the folding protocol.
The stationary state is independent of the initial state.  This decomposition
can also be useful as a way of simulating the Lindblad equation using quantum
gates \cite{mazziotti}.  Using this expression for the NESS in the second
space and equation (\ref{e06221}) for the thermodynamic state, also in the
second space, both states can be compared by looking at the overlap obtained
as the inner product between vector states.
\section{Stationary solutions of the Lindblad equation}
\label{s10023}
As has been shown, both the thermodynamic state as well as the
Lindblad-equation solution can be more conveniently studied in the second
space, not only numerically but also analytically.  Thus, the
question of whether an open quantum system governed by the Lindblad equation
evolves toward a thermal state is explored henceforth by means of inspecting
the solutions of the aforementioned equation.
\subsection{The infinity-temperature solution}
\label{s10026}
{\it
Theorem 1. \\
State
\begin{gather}
|\rho_{it}) = 2^{-N} \bigotimes_{l=1}^N \left [ |0_{2l-1} 0_{2l} ) + x |1_{2l-1}
1_{2l} ) \right ],
\label{e08151}
\end{gather}
is a solution of equation (\ref{e07073}) for Liouvillian (\ref{e07071}) with
irreducible real-Hamiltonian-coefficients, $h_{j,k}$, and bath constants given
by
\begin{gather}
B_{2j-1}^{(n)} = \delta_j^n b_j, \label{e08154} \\ 
B_{2j}^{(n)} = \left( \frac{-1\pm \sqrt{1- x^2}}{x} \right)
B_{2j-1}^{(n)}.
\label{e08152}
\end{gather}
Integer $n$ takes values between $1$ and $N$ and at least one coefficient
$b_n$ must be different from zero.
} \\

Proof.

Let us separate Liouvillian (\ref{e07071}) between the part that contains the
Hamiltonian coefficients $h_{j,k}$ and the part that contains the bath
constants. For a set of bath constants given by (\ref{e08154}) and
(\ref{e08152}) the bath part has the following effect on state (\ref{e08151})
{\scriptsize
\begin{gather}
\sum_{n=1}^N ( 2 B_{2n-1}^{(n)} B_{2n}^{(n)} \tilde{c}_{2n-1}^\dagger
\tilde{c}_{2n}^\dagger + B_{2n-1}^{(n)^2} \tilde{c}_{2n-1}^\dagger
\tilde{c}_{2n-1} + B_{2n}^{(n)^2} \tilde{c}_{2n}^\dagger \tilde{c}_{2n}  )
|\rho_{it}) \nonumber \\ 
=\sum_{n=1}^N [2 B_{2n-1}^{(n)} B_{2n}^{(n)} + x(B_{2n-1}^{(n)^2} +
B_{2n}^{(n)^2})] |\rho_{1,n-1}) |1_{2n-1}1_{2n}) |\rho_{n+1,N}), 
\label{e08153}
\end{gather}
}
replacing
\begin{gather} 
|\rho_{j,k} ) = 2^{j-k}\bigotimes_{l=j}^k \left [ |0_{2l-1} 0_{2l} ) + x |1_{2l-1}
1_{2l} ) \right ].
\label{e08193} 
\end{gather} 
Writing the bath constants in terms of $x$ it can be shown that
\begin{gather} 
2 B_{2n-1}^{(n)} B_{2n}^{(n)} + x(B_{2n-1}^{(n)^2} + B_{2n}^{(n)^2}) = 0.
\label{e07075} 
\end{gather} 
As a consequence, every term of the sum in (\ref{e08153}) equals zero and so
does the whole expression.
Regarding the part of the Liouvillian that involves the Hamiltonian
coefficients, one can check the following result
{\small
\begin{gather} 
\tilde{c}_{2k}^{\dagger} \tilde{c}_{2j-1} |\rho_{it} ) =
x (1-\delta_{j}^{k}) (-1)^{\theta(2k-2j+1)}  
(-1)^{\sum_{q=min(2k,2j-1)}^{max(2k,2j-1)} \tilde{n}_q }  \nonumber \\
| \rho_{1,J-1} ) |0_{2J-1} 1_{2J} ) | \rho_{J+1,K-1} ) |0_{2K-1} 1_{2K} ) 
| \rho_{K+1,N} ),
\label{e08191} 
\end{gather} 
}
taking
\begin{gather} 
J = min(j,k), \hspace{0.25 cm} K=max(j,k).
\label{e08192} 
\end{gather} 
As a consequence, the next identity ensues
\begin{gather} 
h_{k,j} \tilde{c}_{2j}^{\dagger} \tilde{c}_{2k-1} |\rho_{it} ) = -h_{j,k}
\tilde{c}_{2k}^{\dagger} \tilde{c}_{2j-1} |\rho_{it} ). 
\label{e08194} 
\end{gather} 
In a similar way, it can be proved that
\begin{gather} 
h_{k,j} \tilde{c}_{2k-1}^{\dagger} \tilde{c}_{2j} |\rho_{it} ) = -h_{j,k}
\tilde{c}_{2j-1}^{\dagger} \tilde{c}_{2k} |\rho_{it} ). 
\label{e08195} 
\end{gather} 
Consequently, the sum of all elements adds up to zero
\begin{gather} 
\sum_{j=1}^{N} \sum_{k=1}^{N}  h_{j,k} \left (
\tilde{c}_{2 k}^\dagger \tilde{c}_{2 j - 1} + \tilde{c}_{2 j - 1}^\dagger
\tilde{c}_{2 k}\right ) |\rho_{it} )= 0.  
\label{e08196} 
\end{gather} 
Because both bath- and Hamiltonian-parts vanish, equation (\ref{e07073}) is
proved. It is important that at least one coefficient $b_j$ be different from
zero because otherwise all bath constants would vanish and the stationary
state would be indefinite. $\square$ \\

Comparing solution (\ref{e08151}) with respect to the thermodynamic state
(\ref{e06036}) and partition function (\ref{e08041}), it can be seen that an
equivalence between the two is possible insofar as
\begin{gather}
x = \tanh \frac{\alpha_l}{4}.
\label{e08197} 
\end{gather}
This requires all the $\alpha_l$'s to take the same value. Because these
coefficients are related to the system's eigenvalues and parameters by
equation (\ref{e08132}), the only instance a match can be made is when 
\begin{gather}
\beta = 0, \hspace{0.25 cm} \beta \mu = 2 \tanh^{-1} x.
\label{e08198} 
\end{gather}
It becomes in this way noticeable that the solution correspond to a state of
infinity temperature and constant fugacity. As a particular case, all linear
systems with off-diagonal Hamiltonian coefficients and bath operators defined
at a single index location relax toward this type of state.
\subsection{The characteristic solution}
\label{s10027}
{\it
Theorem 2. \\
State
\begin{gather}
|\rho_{th}) = 2^{-N} \bigotimes_{l=1}^{n_b} \bigotimes_{l'=1}^{d_l} \left [
|0_{2L-1} 0_{2L} ) + x_{l} |1_{2L-1} 1_{2L} ) \right ], 
\label{e08211} 
\end{gather}
with
\begin{gather}
L = \sum_{p=1}^l d_{p-1} + l', \nonumber
\end{gather}
is a solution of equation (\ref{e07073}) for Liouvillian (\ref{e07071}) with
real Hamiltonian coefficients that form a block-diagonal matrix of $n_b$
irreducible blocks, each of size $d_l \times d_l$, $l=1,...,n_b$, as follows 
\begin{gather}
h_{J,K} =  y_{j',k'}^k \delta_j^k , \hspace{0.25 cm} J = \sum_{p=1}^j
d_{p-1} + j', \hspace{0.25 cm} K = \sum_{p=1}^k d_{p-1} + k'.
\nonumber
\end{gather}
Both $j$ and $k$ run between $1$ and $n_b$ while $j'$ and $k'$ run from $1$
to $d_j$ and $d_k$ respectively. By definition $d_0 = 0$.
In addition, bath constants must be given by
\begin{gather}
B_{2J-1}^{(n)} = \delta_J^n b_{j'j}, \label{e09061} \\ 
B_{2J}^{(n)} = \left( \frac{-1\pm \sqrt{1- x_j^2}}{x_j} \right)
B_{2J-1}^{(n)}.
\label{e09062}
\end{gather}
Integer $n$ takes values between $1$ and $N$ and there must be at least one
coefficient $b_{j'j}$ different from zero for each value of $j$.
} \\

Proof.

The system can be divided in $n_b$ independent subsystems determined by the
division of blocks in the Hamiltonian. For the $l$'th independent subsystem a
solution analogous to (\ref{e08151}) with bath constants given by
(\ref{e08152}) can be built with $x=x_l$. Accordingly, at least one $b_{j'}$'s
in that subspace must be different from zero. The tensor product of these
solutions can be written as (\ref{e08211}) and constitutes a solution for the
whole system with bath constants defined accordingly on each subspace and the
condition that at least one bath coefficient must be different from zero in
each subspace. $\square$

If one assumes that the Hamiltonian coefficients form a diagonal matrix,
it is possible to establish a parallel between state (\ref{e08211}) and
thermodynamic state (\ref{e06036}) with partition function (\ref{e08041}). 
The equivalence requires
\begin{gather}
x_l = \tanh \frac{\beta (\mu - \epsilon_l)}{2}.
\label{e08213}
\end{gather}
If the Hamiltonian coefficients are not diagonal, it is not possible to
establish an equivalence like (\ref{e08213}) any longer. The incompatibility
can be understood by noticing that if the coefficient matrix is not diagonal,
neither is the corresponding thermodynamic state at finite temperature, while
(\ref{e08211}) turns into a diagonal matrix when shifted to the first space.
That being so, it follows a {\it finite-temperature thermodynamic state of
form  (\ref{e08211}) can only be conceived when the Hamiltonian coefficients
form a completely diagonal matrix}. In such a case all bath coefficients,
which display a local distribution too, must be different from zero in order
to guarantee the uniqueness of the stationary state. This relation between
locality and thermalization has also been observed recently in studies
focusing on conservation laws in open quantum systems \cite{dhar}. 

The characterization above let us establish an operational connection between
driving mechanisms and the distribution of Hamiltonian coefficients.
Specifically, it can be argued that driving terms should manifest with at
least some of their components as non-diagonal coefficients, in such a way
that as a result the stationary state adopt a non-thermal configuration.
As the driving terms become smaller, the Hamiltonian coefficients become more
diagonal and the stationary state should approach the thermodynamic state
obtained from the diagonal part of the coefficients which, along with the bath
constants, relate to the temperature and chemical potential through equation
(\ref{e08213}) in the limit of zero driving.  Moreover, one could proceed by
using the system's parameters in this limit to determine the bath factors
through equation (\ref{e08213}), using fixed values of $\beta$ and $\mu$.
These same bath factors can then be introduced in the Lindblad equation to
find the NESS under the action of driving, so that the state obtained in this
way coincide with a thermodynamic state in the zero-driving limit.  These
results indicate that expression (\ref{e08211}) can potentially correspond
to either thermal- or non-thermal-states, depending on the
distribution of Hamiltonian coefficients, an as such constitutes a powerful
tool to classify the system response, being this the reason why the paradigm
is featured as the characteristic solution.

The scheme can be applied analogously when temperature and chemical potential
are functions of mode index, though the physical justification for this
dependency might be controversial. A different scenario would be to consider
bath interaction localized on a part of the system, for example the edges,
with unequal thermodynamic parameters. Temperature and chemical potential
could be adjusted locally if the Hamiltonian is decoupled, but the part of the
system that is not subject directly to baths would remain indefinite. A
potential solution is to consider the infinity-temperature solution with
different fugacity values on each half of the system. As only one bath
constant must be different from zero on each region, this could be chosen
precisely where the coupling with the baths is supposed to take place.

Both the infinity-temperature- and the characteristic-solutions represent
density matrices that are diagonal in the first space. This stems from the
fact that only bath operators with localized contributions are considered in
each case.  A wider spectrum of solutions can be made available by considering
that unitary transformations preserve the physical significance of the
elements on which they act. On that regard, let us observe that the effect of
an unitary operation $\tilde U$ on equation (\ref{e07073}) can be seen in the
following way
\begin{gather} 
\tilde U \tilde{\mathscr{L}} \tilde U^{-1} \tilde U | \rho_{ss} ) =
\tilde{\mathscr{L}}' | \rho_{ss}' ) = 0.
\end{gather} 
Therefore, $| \rho_{ss}' )$ is a valid density matrix as well as the
stationary state of Liovillian $\tilde{\mathscr{L}}'$.  In this sense, any
density matrix that can be unitarily connected with a solution of the
time-independent Lindblad equation is a solution of a Liouvillian equally
connected to the original Liouvillian. One can thus expand the family of cases
for which the theorems above apply to those instances where Liovillian and
solution can be unitarily connected to complying equivalents. Nevertheless, a
more practical approach is simply to work on a basis where all bath operators
be localized, which should always be possible to do when such operators
commute with each other.
\subsection{Numerical verification}
\label{s10024}
\begin{figure}
\begin{center}
\includegraphics[width=0.35\textwidth,angle=-90]{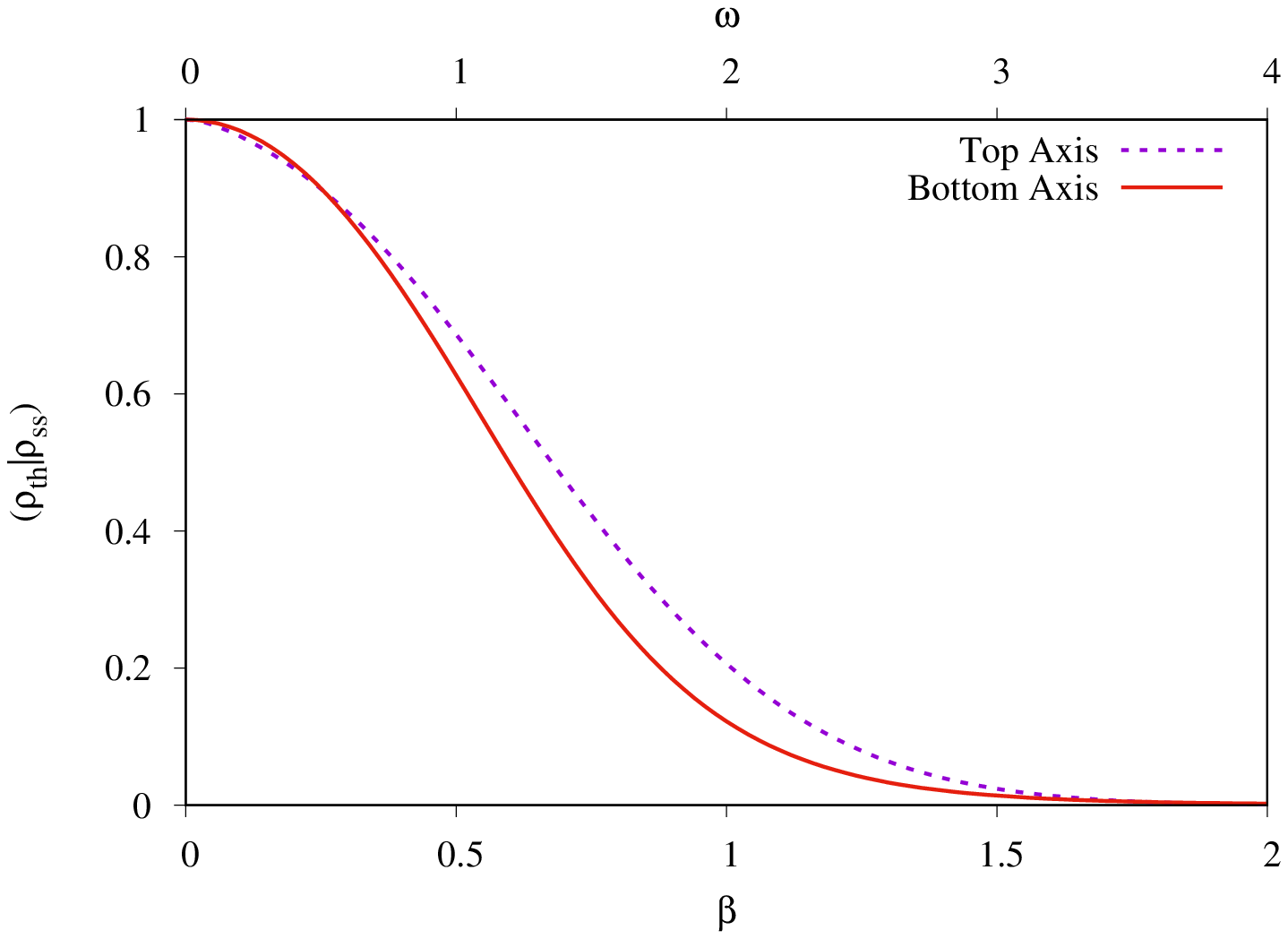}
\caption{Overlap between the stationary- and thermodynamic-states in a system
of size $N=10$. Both states have been normalized in their respective vector
spaces in order to make the maximum possible overlap equal to one. The
numerical values employed can be consulted in the main text. Bottom Axis:
Overlap between the infinity-temperature solution of the Lindblad equation and
the thermodynamic state at inverse temperature $\beta$ and fixed $\beta \mu$.
Both states coincide at $\beta=0$, in accordance with the result that the
stationary solution corresponds to a state of infinity temperature. Top Axis:
Overlap between the thermodynamic state defined by a Hamiltonian with
off-diagonal elements and the Lindblad solution with the diagonal part of the
same Hamiltonian. The overlap is maximum when the off-diagonal terms vanish,
showing that the stationary state obtained from the diagonal Hamiltonian is a
genuine thermodynamic state.}
\label{fig2}
\end{center}
\end{figure}

In order to assess the discussed solutions, let us first consider a set of
Hamiltonian coefficients given by $h_{j,k} = \delta_{j+1}^k + \delta_j^{k+1}$
and bath constants given by (\ref{e08154}) and (\ref{e08152}) with $b_j = 1$,
$\forall j$, and $x=-0.5$. The overlap between the state obtained by
numerically solving the time-independent Lindblad equation (\ref{e07073}) and
the thermodynamic state obtained following the procedure outlined in section
\ref{09063} for a given inverse temperature $\beta$ and $\beta \mu = 2
\tanh^{-1} (-0.5)$ can be seen as the continuous line of figure \ref{fig2}.
Predictably, the overlap between the states reaches its maximum value at
$\beta=0$, showing that the Lindblad solution corresponds indeed to a state of
infinity temperature, in accordance with the analysis following the proof of
the first theorem.

Second, a set of Hamiltonian coefficients given by $h(j,k) = j \delta_j^k +
(1-\delta_j^k)\omega$ is considered. A thermodynamic state for parameters
$\beta=\mu=1$ can be built using the protocol introduced before.  The overlap
between this state and the numerical solution of the Lindblad equation taking
the previous Hamiltonian coefficients with $\omega=0$ and bath constants $x_j
= \tanh \frac{1-j}{2}$ can be seen as the broken line of figure \ref{fig2}.
The overlap reaches its maximum value at $\omega=0$, proving that the
stationary solution with diagonal Hamiltonian-coefficients coincides with the
thermodynamic state of the system defined by such coefficients, exemplifying
an extreme case of the family of solutions established by the second theorem.

\section{Conclusions}
\label{s10025}
The question of thermalization of linear Fermi systems has been addressed from
the perspective of a second operational space and by making use of a reduction
protocol of the quantum state. It is shown how this protocol can be
implemented to encode the thermodynamic state in matrix-product
representation, enabling the comparison with the stationary state of the
Lindblad equation obtained also in matrix-product representation in a related
work. This approach let us, on the one hand, prove two theorems providing a
set of stationary solutions of the Lindblad equation, and on the other hand,
establish how these solutions can in some cases be made to conform to a
thermodynamic state. The first theorem characterizes systems bound to evolve
toward infinity-temperature states. The second theorem provides solutions that
can be fitted to thermal states only in those cases where the Hamiltonian
coefficients adjust to a diagonal matrix. This behavior suggests driving
mechanisms should be incorporated with at least some of their elements as
non-diagonal terms, so that their effect could be differentiated from the
action of the baths which, owning to the special conditions of coupling, are
expected to drift the system toward thermalization. Potential application
areas include, among others, the effect of an electric field on
one-dimensional electrons, the simulation of laser-driving on quantum matter
and the study of fluctuation-dissipation relations in linear systems.
\end{document}